\documentclass[9pt,twocolumn,twoside]{osajnl}
\journal{ol} % Choose journal (ao, aop, josaa, josab, ol, pr)

\setboolean{shortarticle}{true}
\title{All-Digital Stokes Polarimetry with a Digital Micro-mirror Device}
\author[1]{Amogh Manthalkar}
\author[2]{Isaac Nape}
\author[2]{Najmeh TabeBordbar}
\author[3]{Carmelo Rosales-Guzm\'{a}n}
\author[1]{Shanti Bhattacharya}
\author[2]{Andrew Forbes}
\author[2,*]{Angela Dudley}
\affil[1]{Department of Electrical Engineering, Indian Institute of Technology Madras, Chennai – 600036, India}
\affil[2]{School of Physics, University of the Witwatersrand, Private Bag 3, Johannesburg 2050, South Africa}
\affil[3]{Wang Da-Heng Collaborative Innovation Center for Quantum Manipulation and Control, Harbin University of Science and Technology, Harbin 150080, China}

\affil[*]{Corresponding author: angela.dudley@wits.ac.za}

\begin{abstract}
Stokes polarimetry is widely used to extract the polarisation
structure of optical fields, typically from six measurements, although it can be extracted from only four. To measure the required intensities, most approaches are based on optical polarisation components. In this work, we present an all-digital approach that enables a rapid measure of all four intensities without any moving components. Our method employs a Polarisation Grating (PG) to simultaneously project the incoming mode into left- and right-circular polarised states, followed by a polarisation-insensitive Digital Micro-mirror Device (DMD), which digitally introduces a phase retardance for the acquisition of the remaining two polarisation states. We demonstrate how this technique can be applied to measuring the SoP, vectorness and intra-modal phase of optical fields, without any moving components and shows excellent agreement with theory, illustrating fast, real-time polarimetry. 
\end{abstract}

\setboolean{displaycopyright}{true}

\begin{document}
\maketitle
%\section{Introduction}
Polarisation is a remarkable feature of light, which plays a significant role in a variety of areas such as optical fiber communication, radar, metrology, biological microscopy, and 3D movies \cite{PyePolarisation2015, CloudePolarisation2010}. 
%In fact polarisation is the essential component in a range of important quantum optics experiments \cite{aspect1981experimental, aspect1982experimental, bouwmeester1997experimental, kwiat1999ultrabright}.
Therefore, measuring and determining the State of Polarisation (SoP) of light is crucial. In 1852, Sir George Gabriel Stokes showed that it is possible to describe the polarisation state of light by measuring four quantities named Stokes polarisation parameters in which the first parameter describes the total intensity of the light, while the remaining parameters its polarisation state \cite{stokes1851composition}. These four parameters can be used to describe completely polarised, partially polarised and also non-polarised light. 

In order to measure these four Stokes parameters, there are several measurement methods \cite{berry1977measurement, gori1999measuring, schaefer2007measuring, goldstein2016polarized}. One of the most efficient is based on measuring four intensities rather than the usual six, which can be performed by passing a beam through two optical elements - a phase retarder and a polariser, while recording the intensity on a camera. These measured intensities contain sufficient information in order to compute all the Stokes parameters and reconstruct the SoP of the optical field. Conventionally, in order to measure each of these four intensities, one would have to change the angle of the phase retarder and polariser and measure the corresponding intensities one after another. This approach is time-consuming and more prone to error for real-time measurements. To overcome this challenge, methods based on amplitude division have been presented but have other drawbacks such as having to take into account the unbalanced distribution of energy and mismatch in the beam size \cite{azzam1982division}. %\cite{azzam1982division, fridman2010real}. 
Recently, a Digital Micro-mirror Device (DMD) has been used to perform Stokes polarimetry by acting as a grating to divide the incoming mode into four copies of itself for the four polarisation projections \cite{hu2019single}. However, after this step, the Stokes polarimetery is based on the conventional, static approach consisting of phase retarders and polarisers.

Here, we demonstrate an all-digital approach to perform the required polarisation projections for measuring the Stokes parameters. Our scheme uses a DMD \cite{Scholes2019} to realise mode projections onto various polarisation states \cite{hornbeck1996digital}. %\cite{hornbeck1996digital, ren2015tailoring}
This enables us to measure all four intensities by addressing the DMD with appropriate holograms. We tested our technique on vector vortex beams that play a crucial role in a myriad of applications \cite{rubinsztein2016roadmap}. %\cite{rubinsztein2016roadmap, ndagano2017characterizing, ndagano2017creation, rosales2018review}.
In addition, from the Stokes parameters, we can determine the degree of non-separability of the beam \cite{selyem2019basis, mclaren2015measuring}, its SoP \cite{galvez2012poincare} and its intra-modal phase.   

%\section{Concept and Theory}

%\begin{figure}[htbp]
%\centering
%\fbox{\includegraphics[width=\linewidth]{Concept_2}}
%\caption{Conceptual schematic of the all-digital polarisation polarimetry device. DMD: Digital Micro-mirror Device; BS: 50:50 Beam-Splitter; M: Mirror.}
%\label{fig:concept}
%\end{figure}

\begin{figure*}[htbp]
\centering
 \fbox{\includegraphics[width=\textwidth]{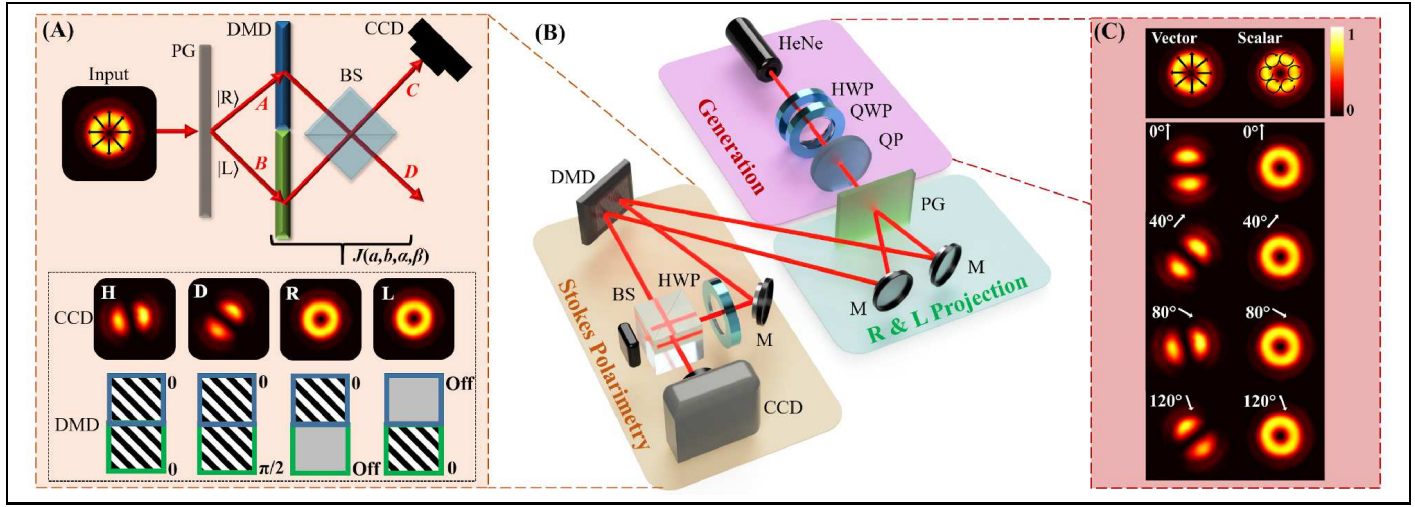}}
\caption{(A) Conceptual schematic of the all-digital polarisation polarimetry device. PG: Polarisation Grating; DMD: Digital Micro-mirror Device; BS: 50:50 Beam-Splitter. The insert contains examples of measured output states with corresponding DMD holograms, where blue denotes the hologram for path \textit{A} and green for path \textit{B}. (B) Experimental setup where the red section demarcates the generation of the mode of interest, while the green and orange sections contain the measurement procedure: first a projection into the left and right polarisation states followed by a projection into the horizontal and diagonal polarisation states. (C) Experimental scalar and radially polarized vector beams. The intensity profiles denote the radially polarized vector beam (left column) and the scalar beam (right column) after passing through a rotating polariser. Polariser angles are marked above each output.}
\label{fig:concept}
\end{figure*}

%\subsection{Stokes Polarimetry}
To illustrate the novelty of our scheme we first explore conventional Stokes polarimetry followed by our approach which exploits path and transverse spatial mode encoding. To demonstrate our technique, we use the well-known cylindrical vector vortex modes, which can be described in standard polar coordinates as, 
\begin{equation}
U(r,\phi,\theta) = \cos(\theta)\exp(-i\ell\phi)\hat{e}_R + \sin(\theta)\exp(i\ell\phi)\exp(i\varphi)\hat{e}_L,
\label{eq:VectorBeam}
\end{equation}
where $\ell$ is the azimuthal index of the beam carrying $\ell\hbar$ quanta of orbital angular momentum (OAM) per photon and $\exp(i\varphi)$ is the intra-modal phase between the orthogonal polarisation states, here left and right. The mode can be varied from purely scalar ($\theta = 0$) to purely vector ($\theta = \pi/4$). 

From a minimum of four intensity measurements, the associated Stokes parameters can be determined,
% \begin{eqnarray}
% S_0 & = & I_R + I_L,\,\,\,\,\,\,\nonumber\\
% S_1 & = &  2I_H - S_0,\nonumber\\
% S_2 & = & 2I_D - S_0,\nonumber\\  
% S_3 & = & I_R - I_L, \label{eq:Stokes}
% \end{eqnarray}
\begin{eqnarray}
S_0 & = & I_R\, +\, I_L,\,\,\,\,\,\,\,\,S_1\,\,\, =\,\,\,  2I_H - S_0,\nonumber\\
S_2 & = & 2I_D - S_0,\,\,\,\,\,\,S_3\,\,\, =\,\,\,  I_R\, - I_L\,, \label{eq:Stokes}  
\end{eqnarray}
where $I_H$, $I_D$, $I_L$ and $I_R$ represent the two-dimensional intensity profiles of the horizontal, diagonal, left- and right-circular polarisation components, respectively. 

The Stokes parameters enable not only the spatial reconstruction of the SoP of the optical field of interest (Eq. \ref{eq:VectorBeam}), but the Vector Quality Factor (VQF) \cite{selyem2019basis, mclaren2015measuring} can also be determined. 
\begin{equation}
VQF = \sqrt{1 - \frac{S_1^2}{S_0^2} - \frac{S_2^2}{S_0^2} - \frac{S_3^2}{S_0^2}},
\label{eq:Vectorness}
\end{equation}
as well as the intra-modal phase \cite{goldstein2016polarized}, 
\begin{equation}
\varphi(r,\phi) = \frac{1}{2}\arctan(\frac{S_3(r,\phi)}{S_2(r,\phi)}).
\label{eq:Phase}
\end{equation}
In order to perform Stokes polarimetry via the four intensity measurements listed in Eq. \ref{eq:Stokes}, we need to perform projections into the horizontal, diagonal, left- and right-circular polarisation states. Conventionally, this is achieved with static phase retarders and polarisers. In the next section we show how these measurements can be executed using path encoding and interference.

%\subsection{A Static and Digital Approach}
In the previous section, the role of the polariser was to enable the interference of the spatial modes, marked by the orthogonal polarisations. For example, a vector mode $\Psi(r, \phi) = \frac{1}{\sqrt{2}} \left[ \psi_R(r, \phi) \hat{e}_R + \psi_L(r, \phi) \hat{e}_L \right]$ projected onto an arbitrary polarisation state [i.e,  $ \hat{a} \left( \alpha \right) = \frac{1}{\sqrt{2}} \left(u \hat{e}_R + v e^{i\alpha} \hat{e}_L \right) $ where $\alpha$ (phase), \textit{u} and \textit{v} (amplitudes) are determined by the orientation of the polariser] results in a field expressed as $ \left( u\psi_R(r, \phi)  + ve^{ -i \alpha} \psi_L(r, \phi) \right) \hat{a}(\alpha) $. Therefore, any projection with a linear polariser controls the phase and amplitude between the spatial modes that where initially marked by unique polarisation states.  In our scheme, we emulate this by using path decomposition of the orthogonal circular polarisation in combination with digital phase encoding and path interference using a beam-splitter. We have illustrated this in Fig. \ref{fig:concept} (A).

For the polarisation decomposition, we use a geometric phase Polarisation Grating (PG), which performs a decomposition of left- and right-circular polarisations, onto orthogonal paths \textit{A} and \textit{B},  respectively, and hence we can extract $I_L$ and $I_R$ in a single measurement. At this point in the setup, the orthogonal states of the field of interest [$\Psi^{'}(r, \phi)$], namely $\psi_R(r, \phi)$ and $\psi_L(r, \phi)$, are mapped to two unique paths, namely \textit{A} and \textit{B}, respectively, which is represented as a vector as follows $\Psi^{'}(r,\phi)=\big(\begin{smallmatrix}
\psi_R(r, \phi)\\
\psi_L(r, \phi)
\end{smallmatrix}\big).$ Here, the first and second entries map the fields in path \textit{A} and \textit{B}, respectively [as illustrated in Fig. \ref{fig:concept} (A)]. Each path (or orthogonal state, $\psi_R(r, \phi)$ and $\psi_L(r, \phi)$) is directed to one half of a DMD, where a relative phase and amplitude is encoded between $\psi_R(r, \phi)$ and $\psi_L(r, \phi)$, represented by the following matrix  $\big(\begin{smallmatrix}
a\exp(i\alpha) & 0\\
0 & b\exp(i\beta)
\end{smallmatrix}\big).$ 
The two paths are then combined via a 50:50 Beam-Splitter (BS), which has an unitary matrix given as $\frac{1}{\sqrt{2}}\big(\begin{smallmatrix}
1 & 1\\
1 & -1
\end{smallmatrix}\big)$. The Jones matrix for the final transformation is summarised as
 \begin{equation}
 J(a, b,\alpha,\beta) = \frac{1}{\sqrt{2}}
  \begin{pmatrix}
    a\exp(i\alpha) & b\exp(i\beta) \\
    a\exp(i\alpha) & -b\exp(i\beta)
  \end{pmatrix}, \label{eq:Jones}
 \end{equation}
where the amplitudes ($a$ and $b$) and phases ($\alpha$ and $\beta$) are controlled on the DMD. Applying this transformation (Eq. \ref{eq:Jones}) to our field of interest [$\Psi^{'}(r, \phi)$], results in the following state
\begin{equation}
 J(a, b,\alpha,\beta) \Psi^{'}(\cdot)=
  \frac{1}{2}\begin{pmatrix}
    a\exp(i\alpha) \psi_R(r, \phi) + b\exp(i\beta)\psi_L(r, \phi) \\
    a\exp(i\alpha) \psi_R(r, \phi) -b\exp(i\beta) \psi_L(r, \phi)
  \end{pmatrix}. \label{eq:OutputMode}
 \end{equation}
\noindent
where the top (bottom) entry denotes path \textit{C} (\textit{D}) located after the BS in Fig. \ref{fig:concept} (A). By substituting the parameters (provided in Table \ref{table:1}) into the final state located at the CCD in path \textit{C} (i.e. the top entry of the vector in Eq. \ref{eq:OutputMode}, one can obtain the four required states, namely horizontal, diagonal, right and left.

The settings for the DMD are presented in Table \ref{table:1}.
\begin{table}[h!]
\centering
\caption{Parameter settings in order to project onto the corresponding polarisation states using the Jones matrix $J(a, b, \alpha, \beta)$.}

$\begin{array}{c|cccc |cccc}
 & a & b & \alpha & \beta \,\,\,\,\,\,\,\,\,\,\,\,\,\,\,\,\,\,\,\,\,\,\,\, & a & b & \alpha & \beta\\
\hline
I_H & 1 & 1 & 0 & 0 \,\,\,\,\,\,\,\,\,\,\,\,\,\,\,\,\,\,\,\,\,\,\,\, I_D & 1 & 1 & 0 & -\frac{\pi}{2}\\
I_R & 0 & 1 & 0 & 0 \,\,\,\,\,\,\,\,\,\,\,\,\,\,\,\,\,\,\,\,\,\,\,\, I_L &  1 & 0 & 0 & 0 \\
\end{array}
\label{table:1}
$\end{table}

%\begin{array}{c|cccc}
% & a & b & \alpha & \beta\\
%\hline
%H & 1 & 1 & 0 & 0 \\
%D & 1 & 1 & 0 & -\frac{\pi}{2}\\
%R & 0 & 1 & 0 & 0 \\
%L &  1 & 0 & 0 & 0 \\
%\end{array}
%\label{table:1}
%\end{table}

%\section{Experimental Setup}

%\begin{figure}[htbp]
%\centering
%\fbox{\includegraphics[width=\linewidth]{SetUp.png}}
%\caption{Experimental setup of the all-digital polarisation polarimetry device. The red section demarcates the generation of the mode of interest, while the green and orange sections contain the measurement procedure: first a projection into the left and right states followed by a projection into the horizontal and diagonal states. HeNe: Helium-Neon laser; HW: Half-Wave Plate; QW: Quarter-Wave Plate; PG: Polarisation Grating; CCD: Charged-Coupled Device.} 
%%HeNe: Helium-Neon laser; HW: Half-Wave Plate; QW: Quarter-Wave Plate; PG: Polarisation Grating; M: Mirror; DMD: Digital Micro-mirror Device; BS: 50:50 Beam-Splitter; CCD: Charged-Coupled Device.}
%\label{fig:setup}
%\end{figure}

The experimental setup for the all-digital Stokes polarimetry device is depicted in Fig. \ref{fig:concept} (B). Here the setup was classified into three sections: generation of the mode of interest (scalar, vector or a weighted superposition), projection into the left- and right-circular polarisation states, followed by Stokes polarimetry, executed via a DMD and a BS. A horizontally polarised Gaussian beam (Helium-Neon, $\lambda$ = 632 nm) was converted into a radial vector vortex beam with the use of a static q-plate ($q = 1/2$), a Half Wave-Plate (HWP) and Quarter Wave-Plate (QWP). Once the mode of interest was created, a PG was used to separate the left and right-circular polarisation states, which were subsequently directed onto two halves of a DMD.

To record the left-circular polarisation state ($I_L$), the half of the DMD associated to the incident left-circular polarisation state (i.e. path \textit{A}) was in the 'on state'. While the other half of the DMD, associated to the incident right-circular polarisation state (i.e. path \textit{B}), was in the 'off state'. The state of each half of the DMD was then reversed in order to record the right-circular polarisation state ($I_R$) [as illustrated in the insert of Fig. \ref{fig:concept} (A)]. Next the two halves of the DMD were encoded with holograms that produce a relative phase difference of 0 for the detection of the horizontally polarised state ($I_H$), and a phase difference of $\frac{\pi}{2}$ for the diagonal polarisation state ($I_D$), when both beams interfered on the CCD [as illustrated in the insert of Fig. \ref{fig:concept} (A)].

The frame rate of the DMD allowed for a fast ($\sim$9500 1-bit frames per second), real-time measurement of the Stokes parameters. The holograms were generated using the Lee holography method, by passing the desired intensity and phase functions through a non-linear limiter that acts like a binary filter \cite{lee1974binary}.
%\cite{lee1974binary,lee1979binary}. 
%The setup is a 4-\textit{f} setup with an aperture in the Fourier plane to select a particular order of interest. 

The equation used to encode the binary hologram is
\begin{eqnarray*}
h\left( x,y\right)&=&\frac{mod_{2\pi}\left[\phi\left(x,y\right)+2\pi\left(k\,x+k\,y\right)\right]}{2\pi}-0.5\\
&\leq& \frac{\sin^{-1}{\left(\vert E\left( x,y\right)\vert\right)}}{\pi/2}
\end{eqnarray*}

In this equation, $h(x,y)$ is the binary value of each point on the DMD, $E(x,y)$ and $\phi(x,y)$ are the normalized amplitude and phase at each point in the desired pattern, $k$ is the carrier frequency, which determines the fringe width. %This determines how far different orders will be separated in the Fourier plane. This particular parameter presents an interesting trade-off between phase resolution and selectivity. 
Since the phase functions that we generated are constant global phases, the holograms look like binary gratings with relative shifts in position for different phase values [Fig. \ref{fig:concept} (A)].

Since the paths that the left- and right-circularly polarized beams traversed after reflection from the DMD were different, a calibration experiment was performed to verify the phase that each beam acquires from the DMD. For this, the DMD was programmed with a constant hologram on one half and a sequence of holograms for a constant global phase were displayed. The hologram for which an interferogram corresponding to $I_H$ (two horizontal lobes) was captured was then considered to be a relative phase of 0 between both beams at the CCD. A hologram corresponding to a phase difference of $\frac{\pi}{2}$ was programmed to obtain $I_D$. The measured intensity profiles ($I_H$, $I_D$, $I_L$ and $I_R$) were used to calculate the associated Stokes parameters and subsequently the VQF of the mode, its SoP and intra-modal phase (all presented in the following section, pertaining to the \textit{Results}).

%\section{Results}

The concept of the all-digital Stokes polarimetry measurement was verified with both scalar and radial vector vortex modes, whose intensity profiles are depicted in Fig. \ref{fig:concept} (C). Unlike the scalar mode, the vector mode possesses a spatial dependency on the state of polarisation, as illustrated in the series of inhomogeneous intensity profiles recorded after a rotating polariser [Fig. \ref{fig:concept} (C) left column]. When the polariser is orientated to transmit vertically polarised light, two lobes are orientated along the vertical axis [Fig. \ref{fig:concept} (C) left column] revealing the radial nature of the vector mode's polarisation. 

%\begin{figure}[htbp]
%\centering
%\fbox{\includegraphics[width=\linewidth]{Scalar_vs_Vector}}
%\caption{Experimental scalar and radially polarized vector beams. The intensity profiles denote the scalar beam (top row) and the radially polarized vector beam (bottom row) after passing through a rotating polariser. Polariser angles are marked above each output.}
%%Accompanying videos of the intensity profiles after passing through a rotating polariser can be viewed in Visualization 1 and Visualization 2.}
%\label{fig:ScalarVSVector}
%\end{figure}

\begin{figure}[htbp]
\centering
\fbox{\includegraphics[width=\linewidth]{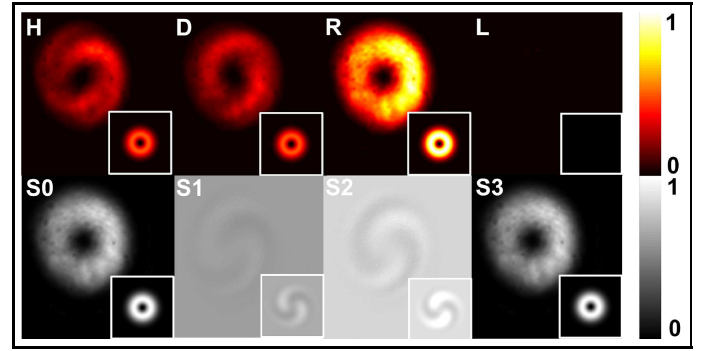}}
\caption{Stokes polarimetry for a scalar mode. The measured intensity profiles (top row) of the Stokes projections with calculated Stokes parameters (bottom row). Theoretical predictions are given as inserts.}
%\caption{Stokes polarimetry for a scalar mode. The measured intensity profiles of the Stokes projections for the scalar beam, labeled H: Horizontal; D: Diagonal; R: Right and L: Left, with calculated Stokes parameters (S0, S1, S2 and S3). Theoretical predictions are given as inserts.}
\label{fig:scalarPOVM}
\end{figure}

\begin{figure}[htbp]
\centering
\fbox{\includegraphics[width=\linewidth]{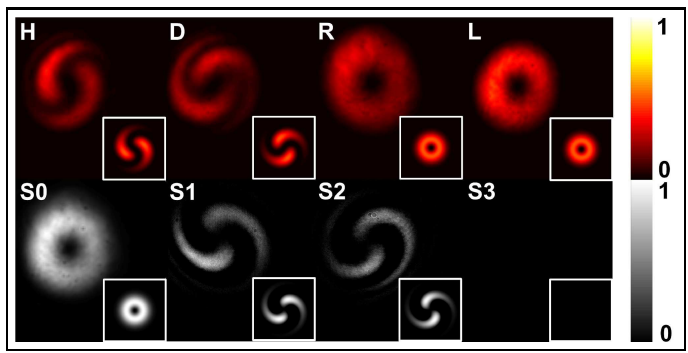}}
\caption{Stokes polarimetry for a radial vector mode. The measured intensity profiles (top row) of the Stokes projections with calculated Stokes parameters (bottom row). Theoretical predictions are given as inserts.}
%\caption{Stokes polarimetry for a radial vector mode. The measured intensity profiles of the Stokes projections for the radial vector beam, labeled H: Horizontal; D: Diagonal; R: Right and L: Left, with calculated Stokes parameters (S0, S1, S2 and S3). Theoretical predictions are given as inserts.}
\label{fig:vectorPOVM}
\end{figure}

The experimental Stokes measurements for the scalar and radially polarised vector beam are displayed in Figs \ref{fig:scalarPOVM} and  \ref{fig:vectorPOVM}, respectively. The top rows contain the intensity profiles of the four intensity profiles (namely, $I_H$, $I_D$, $I_L$ and $I_R$) needed to determine the Stokes parameters, presented in the corresponding bottom rows. In the scalar case (Fig. \ref{fig:scalarPOVM}) it is evident that only one of the circular polarisation states exists (in this case the right-handed state) which is characteristic of these modes, unlike in the vector case (Fig. \ref{fig:vectorPOVM}) where both the left- and right-handed polarisation states are present. The corresponding Stokes parameters determined via Eq. \ref{eq:Stokes} are presented in the bottom rows of Figs \ref{fig:scalarPOVM} and \ref{fig:vectorPOVM}, for the scalar and vector case, respectively.  In both cases there is very good agreement with theoretical predictions given by the accompanying inserts. The spiraling structure of the two-lobe intensity profile (which manifests in Stokes parameters, S1 and S2) is due to the inherent curvature induced in the modes propagating through the experimental setup [Fig. \ref{fig:concept} (B)].

\begin{figure}[htbp]
\centering
\fbox{\includegraphics[width=\linewidth]{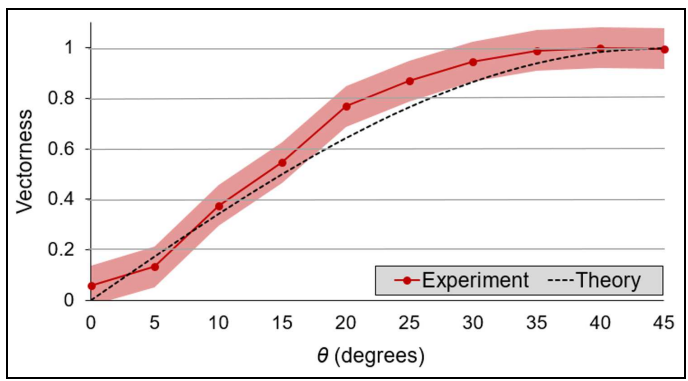}}
\caption{Plot of the measured vectorness (red data points) versus $\theta$ (Eq. \ref{eq:VectorBeam}). The black, dashed curve denotes the theoretical prediction and the red band the measured error.}
\label{fig:vectorness}
\end{figure}

\begin{figure}[htbp]
\centering
\fbox{\includegraphics[width=\linewidth]{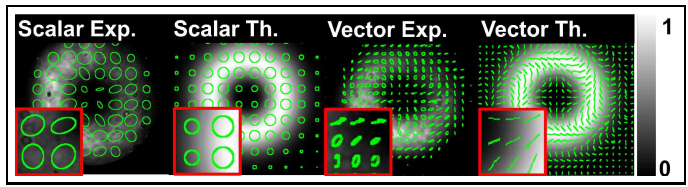}}
%\fbox{\includegraphics[width=5.3cm]{PolarisationPlot}}
\caption{Reconstructed polarisation structure for scalar and vector modes. Theoretical predictions are given alongside. An enhanced view of the polarisation vectors is given as an insert.}
\label{fig:polarisation}
\end{figure} 

\begin{figure}[htbp]
\centering
\fbox{\includegraphics[width=\linewidth]{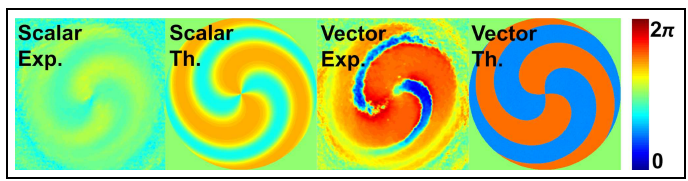}}
\caption{The extracted phase between two orthogonal polarisation states (namely, left and right) for scalar and radial vector modes. Theoretical predictions are given alongside.}
\label{fig:phase}
\end{figure}

The results presented above represent a purely scalar or purely vector case, however similar measurements can be performed on intermediate modes to determine their associated vectorness (as calculated via  Eq. \ref{eq:Vectorness} and shown in Fig. \ref{fig:vectorness}). Here the QWP [depicted in the 'Generation' section of the experimental setup, Fig. \ref{fig:concept} (B)] was rotated so as to create modes, whose polarisation profile varied from homogeneous (scalar) to inhomogeneous (vector). Their associated vectorness was measured via the all-digital Stokes polarimetry technique. Visually, there is high-fidelity between the measured and theoretically expected vectorness values of the generated modes. 

In addition to determining the SoP of the generated scalar and vector modes (Fig. \ref{fig:polarisation}), the phase relationship between orthogonal polarisation states can also be measured, thereby reconstructing the associated wavefront in real-time (Fig. \ref{fig:phase}). The reconstructed polarisation structure of the experimentally generated scalar mode is characteristic of  right-circularly polarised light (first image of Fig. \ref{fig:polarisation}) and is in good agreement with the simulated result (second image of Fig. \ref{fig:polarisation}). While the SoP of the generated vector mode depicts a radial structure (third and last image of Fig. \ref{fig:polarisation}). The phase (or wavefront) between the orthogonal polarisation states calculated via Eq. \ref{eq:Phase}, for both the scalar and vector case, are experimentally reconstructed and displayed in Fig.  \ref{fig:phase}, illustrating the spherical curvature which is incurred by the propagation of the modes, which agrees visually with the theoretical simulated results.  

%\section{Conclusion}

Here we outline an approach to reconstruct the SoP, as well as to measure vectorness and intra-modal phase of an optical field digitally with no moving parts and in real-time by implementing an inhomogeneous polarisation optic (namely, a PG) and a polarisation insensitive DMD. We tested our technique on modes of varying vectorness and obtained high-fidelity between the experimentally reconstructed SoP, vectorness and intra-modal phase with their theoretical counterparts. This technique will prove useful for the development of  robust, compact polarimetery sensors for fields such as optical metrology. 

\medskip

\noindent\textbf{Funding.} Work funded in India through Grant No. DST/IMRCD/BRICS/Pilotcall1/OPTIMODE/2017. 

\medskip

\noindent\textbf{Disclosures.} The authors declare no conflicts of interest.

% Bibliography
\bibliography{sample}

% Full bibliography added automatically for Optics Letters submissions; the following line will simply be ignored if submitting to other journals.
% Note that this extra page will not count against page length
\bibliographyfullrefs{sample}

%Manual citation list
%\begin{thebibliography}{1}
%\bibitem{Zhang:14}
%Y.~Zhang, S.~Qiao, L.~Sun, Q.~W. Shi, W.~Huang, %L.~Li, and Z.~Yang,
 % \enquote{Photoinduced active terahertz metamaterials with nanostructured
  %vanadium dioxide film deposited by sol-gel method,} Opt. Express \textbf{22},
  %11070--11078 (2014).
%\end{thebibliography}

% Please include bios and photos of all authors for aop articles
\ifthenelse{\equal{\journalref}{aop}}{%
\section*{Author Biographies}
\begingroup
\setlength\intextsep{0pt}
\begin{minipage}[t][6.3cm][t]{1.0\textwidth} % Adjust height [6.3cm] as required for separation of bio photos.
  \begin{wrapfigure}{L}{0.25\textwidth}
    \includegraphics[width=0.25\textwidth]{john_smith.eps}
  \end{wrapfigure}
  \noindent
  {\bfseries John Smith} received his BSc (Mathematics) in 2000 from The University of Maryland. His research interests include lasers and optics.
\end{minipage}
\begin{minipage}{1.0\textwidth}
  \begin{wrapfigure}{L}{0.25\textwidth}
    \includegraphics[width=0.25\textwidth]{alice_smith.eps}
  \end{wrapfigure}
  \noindent
  {\bfseries Alice Smith} also received her BSc (Mathematics) in 2000 from The University of Maryland. Her research interests also include lasers and optics.
\end{minipage}
\endgroup
}{}

\end{document}